\documentclass[pra,aps,showpacs,floatfix,twocolumn,nofootinbib,superscriptaddress]{revtex4-1}

\usepackage{amsmath}
\usepackage{amsfonts}
\usepackage{amssymb}
\usepackage{graphicx}
\graphicspath{{figures/}}
\usepackage{xcolor} 

\usepackage{physics}
\usepackage{import}
\usepackage{hyperref}  
\usepackage{dcolumn}
\usepackage{bm}
\usepackage{booktabs} 
\def\redmel#1#2#3{\langle#1\| #2 \| #3 \rangle}

\begin{document}
\title{Measurement of the reduced dipole matrix element in Ba$^+$} 

\author{N. Jayjong}
\author{M. D. K. Lee}%
\affiliation{Centre for Quantum Technologies, National University of Singapore, 3 Science Drive 2, 117543 Singapore}%
\author{K. J. Arnold}%
\affiliation{Centre for Quantum Technologies, National University of Singapore, 3 Science Drive 2, 117543 Singapore}%
\affiliation{Temasek Laboratories, National University of Singapore, 5A Engineering Drive 1, 117411 Singapore
}%
\author{M. D. Barrett}
 \email{phybmd@nus.edu.sg}
\affiliation{Centre for Quantum Technologies, National University of Singapore, 3 Science Drive 2, 117543 Singapore}%
\affiliation{
Department of Physics, National University of Singapore, 2 Science Drive 3, 117551 Singapore
}%

\begin{abstract}
We present a high-precision measurement of the reduced electric-dipole matrix element $\redmel{P_{1/2}}{r}{S_{1/2}}$ in $^{138}\mathrm{Ba}^+$. By comparing off-resonant scattering rates with dispersive Stark-shift measurements, we determine the matrix element to be $3.322\,7(12)$, corresponding to a $P_{1/2}$ excited-state radiative lifetime of $7.866\,3(56)$~ns. Combining our experimental results with a prior model for the differential scalar polarizability $\Delta\alpha_0(\omega)$, we extract the static value $\Delta\alpha_0(0) = -73.09(12)$~a.u. This determination directly improves the evaluation of the blackbody radiation (BBR) shift, minimizing a prominent systematic limitation in room-temperature $\mathrm{Ba}^+$ optical clock error budgets. These results provide a stringent benchmark for atomic-structure calculations and advance high-precision applications in quantum metrology and tests of fundamental physics.
\end{abstract}

\maketitle


\section{Introduction}
\label{sec:Introduction}
Singly ionized barium ($\mathrm{Ba}^+$) is a premier system for precision spectroscopy and the validation of relativistic atomic structure calculations. Featuring a single valence electron outside a closed shell, its electronic structure allows highly accurate, \textit{ab initio} many-body theoretical treatments while remaining sensitive to core electron-correlation and relativistic effects \cite{safronova2003,guet1991relativistic,gopakumar2002electric,iskrenova2008theoretical,dzuba2011calculation}. Beyond testing atomic theory, $\mathrm{Ba}^+$ is a vital platform for optical frequency standards \cite{sherman2005progress,arnold2020precision}, quantum information processing \cite{dietrich2010hyperfine,inlek2017multispecies,hucul2017spectroscopy,bruzewicz2019dual,low2025control}, and low-energy tests of the Standard Model, such as atomic parity nonconservation \cite{fortson1993,dzuba2011calculation}. It is also valuable for modeling clock transition polarizabilities \cite{jayjong2026zero}. Ultimately, the precision limits of these applications depend on minimizing uncertainties in the underlying electric-dipole matrix elements.

One of the most precise experimental determinations of electric-dipole transition strengths was made in $\mathrm{Ba}^+$ from analyses of Rydberg fine-structure measurements in neutral and singly ionized barium, from which the relevant dipole matrix elements were extracted \cite{woods2009dipole, woods2010dipole}. Later, a complementary high-precision method was demonstrated in $\mathrm{Ca}^+$, where a dipole matrix element was measured using a single trapped ion by combining ac-Stark shift measurements with photon-scattering rate measurements \cite{hettrich2015measurement}. Given the close similarities in the electronic structure of $\mathrm{Ca}^+$ and $\mathrm{Ba}^+$, this methodology can be extended to the $\mathrm{Ba}^+$ system with only minor modifications.  As we will show, the larger branching fraction of decays from $P_{1/2}$ to $D_{3/2}$ allows a more straightforward implementation in which many systematics are more heavily suppressed.

In this work, we present a high-precision measurement of the reduced electric-dipole matrix element $\redmel{P_{1/2}}{r}{S_{1/2}}$ in $^{138}\mathrm{Ba}^+$ using a combined analysis of off-resonant scattering rates and Stark-shifts induced by a probe detuned from the $S_{1/2}-P_{1/2}$ transition.  Our measured value of $3.322\,7(12)\,ea_0$ for $\redmel{P_{1/2}}{r}{S_{1/2}}$ provides a rigorous, high-precision challenge to existing experimental benchmarks \cite{jayjong2026zero,woods2010dipole,kastberg1993measurements,davidson1992oscillator,gallagher1967oscillator} and represents a stringent test for theoretical calculations \cite{roberts2023electric,sahoo2009light,dzuba2011calculation,iskrenova2008theoretical,gopakumar2002electric,guet1991relativistic}.  In conjunction with previous work \cite{arnold2019measurements,zhang2020branching,jayjong2026zero}, this single matrix element allows updates to lifetimes of the $P_{1/2}$ and $P_{3/2}$ levels, the associated matrix elements, and the differential scalar polarizability $\Delta\alpha_0(\omega)$ model obtained in \cite{jayjong2026zero}.

\section{Methodology}
\label{Methodology}
The reduced matrix element (RME) $\mu=\redmel{P_{1/2}}{r}{S_{1/2}}$, defined in the Wigner--Eckart convention $\langle J m|r_q|J'm'\rangle=\langle J\|r\|J'\rangle\,\langle J'm'\,1q|Jm\rangle$ and quoted throughout in atomic units, is determined by comparing scattering rates out of $S_{1/2}$ and Stark shifts from a probe beam detuned by $\Delta$ from resonance with the $S_{1/2}-P_{1/2}$ transition as first demonstrated in \cite{hettrich2015measurement}.  Neglecting Zeeman shifts and other transitions, the scalar Stark-shift of the $S_{1/2}$ transition within the rotating wave approximation (RWA) is given by
\begin{equation}
\label{Eq:BasicSS}
\Delta_s=\frac{\Omega_0^2}{24 \Delta},
\end{equation}
where $\Omega_0=\redmel{P_{1/2}}{r}{S_{1/2}}E_0/\hbar$.  Under the same conditions, the scattering rate out of $S_{1/2}$ for linearly polarized light is given by
\begin{equation}
\label{gammas}
\gamma_s= p \Gamma  \frac{\Omega_0^2}{24\Delta^2}= \gamma_{3/2}\frac{\Omega_0^2}{24\Delta^2}=\frac{p \gamma_{1/2}}{1-p} \frac{\Omega_0^2}{24\Delta^2},
\end{equation}
where $p$ is the branching fraction for decays from $P_{1/2}$ to $D_{3/2}$, $\gamma_{1/2}$ and $\gamma_{3/2}$ are the decay rates from $P_{1/2}$ to $S_{1/2}$  and $D_{3/2}$, respectively, and $\Gamma=\gamma_{1/2}+\gamma_{3/2}$.  We then have
\begin{equation}
\label{Eq:Basic}
\frac{\gamma_{1/2}}{\Delta}=\frac{1-p}{p}\frac{\gamma_s}{\Delta_s},
\end{equation}
which is independent of the laser intensity.  Since $p$ has been measured with high accuracy \cite{arnold2019measurements}, $\gamma_{1/2}$ is readily determined from the measured ratio $\gamma_s/\Delta_s$, from which $\mu^2$ is obtained via
\begin{equation}
\label{Eq:Basic2}
\gamma_{1/2} = \frac{2}{3}\,\alpha\,c\,k_{1/2}^3\,\mu^2,
\end{equation}
where $\alpha$ is the fine-structure constant, $c$ is the speed of light, and $k_{1/2}$ is the wavenumber corresponding to the $P_{1/2}$ decay to $S_{1/2}$.

Modifications to Eq.~\ref{Eq:Basic} arise from corrections to the RWA and coupling to other transitions, which include contributions to $\Delta_s$ from $D_{5/2}$ as Stark shifts are measured on the $S_{1/2}-D_{5/2}$ clock transition.  Leading order corrections arise from contributions to $\Delta_s$ due to the scaling with detuning.  However these can be mitigated by comparing results at equal and opposite detunings.  Leading order corrections to Eq.~\ref{Eq:BasicSS} do not change with the sign of the detuning and can thus be eliminated differentially.  Next order corrections then scale similarly with corrections to the scattering rate.   For detunings used in this work, fractional corrections to Eq.~\ref{Eq:Basic} can be kept below $10^{-5}$ when comparing results at equal and opposite detunings.

Consideration must also be given to the polarization components of the probe.  An imbalance in circular components of the polarization contributes to the Stark shift via the vector polarizability, but this is cancelled by averaging the clock transition over a Zeeman pair.  However, the imbalance can impair the scattering rate, which is evident in the extreme limit of optical pumping to a single Zeeman state.  In the most general case of a probe propagating along $\hat{\mathbf{r}}$ with a polarization $\hat{\boldsymbol{\epsilon}}=\cos\beta\, e^{i \chi/2}\hat{\boldsymbol{\theta}}+\sin\beta\, e^{-i \chi/2}\hat{\boldsymbol{\phi}}$, the vector polarizability gives rise to an effective magnetic field proportional to
\begin{equation}
i (\hat{\boldsymbol{\epsilon}}^*\times \hat{\boldsymbol{\epsilon}})=\sin(2\beta) \sin{\chi}\,\hat{\mathbf{r}},
\end{equation}
where $\hat{\mathbf{r}}$, $\hat{\boldsymbol{\theta}}$, and $\hat{\boldsymbol{\phi}}$ are the unit vectors in spherical polars with the $z$-axis determined by the applied static magnetic field, which is taken as the quantization axis.   The effective magnetic field from the probe is thus directed along the propagation direction with a magnitude proportional to the phase difference between two orthogonal projections of the polarization.  Within the RWA and ignoring other transitions, the Stark shifts may be expressed as
\begin{equation}
\frac{\Omega_0^2}{24 \Delta}\left(1-2(|u_{-1}|^2-|u_{1}|^2)m\right),
\end{equation}
where $u_q$ are the spherical components of the polarization with the convention that $\hat{\boldsymbol{\epsilon}}=\sum_q (-1)^q u_{-q}\hat{\mathbf{e}}_q$.  The imbalance in $u_{\pm1}$ given by
\begin{equation}
\label{Eq:imbalance}
\epsilon=|u_{-1}|^2-|u_{1}|^2=-\sin(2\beta) \sin{\chi}\cos\theta,
\end{equation}
can then be determined by the measured Zeeman shift induced by the probe beam.  This assumes the vector shift is much less than the Zeeman splittings from the applied static magnetic field, which is true by design.  In practice we tune the polarization to minimize the vector shift and hence suppress the departure from linear polarization.

Zeeman shifts and departures from linear polarization modify the scattering rates when the atom is in $\ket{S_{1/2}, m=\pm \tfrac{1}{2}}$ from the nominal value of $\gamma_s$ given by Eq.~\ref{gammas}, which results in a deviation of $S_{1/2}$ population decay from a simple exponential decay.  Ignoring coherences, population dynamics can be described by rate equations for the populations $P_\pm$ of $\ket{S_{1/2}, m=\pm \tfrac{1}{2}}$, which are best expressed in terms of the total ground state population $P=P_++P_-$ and the differential population $\delta P=P_+-P_-$.  To lowest order in $\epsilon$, as given by Eq.~\ref{Eq:imbalance}, and $x=\mu_B B/(\hbar\Delta)$, it is readily found that
\begin{equation}
\label{approx2}
\begin{pmatrix}\dot{P}\\ \dot{\delta P}\end{pmatrix}=-\gamma_s \begin{pmatrix} 1 & -\epsilon_1 \\ -\epsilon_2 & k \end{pmatrix}\begin{pmatrix}P \\ \delta P\end{pmatrix},
\end{equation}
where
\begin{subequations}
\begin{align}
\epsilon_1&=\epsilon+g_+ x(1-|u_0|^2)+g_- x |u_0|^2,\\
\epsilon_2&=k_1\epsilon+k_1 g_+ x(1-|u_0|^2)+k_2 g_- x |u_0|^2,\\
k&=k_1(1-|u_0|^2)+k_2 |u_0|^2,
\end{align}
\end{subequations}
with the parameters $k_i$ given by
\[
k_1=\frac{2+p}{3p},\quad k_2=\frac{4-p}{3p},
\]
and $g_\pm = g_S\pm g_P$ being the sum and difference of the Land\'{e} g-factors $g_S$ and $g_P$ for the $S_{1/2}$ and $P_{1/2}$ levels respectively.  Starting with all the population in the $S_{1/2}$ level, i.e. $P(0)=1$, with a population differential $\delta P(0)=\delta$, the solution is approximately
\begin{equation}
\label{Eq:scattering}
P(t)=e^{-\gamma_s t}+\tfrac{\delta \epsilon_1}{k-1}(e^{-\gamma_st}-e^{-k\gamma_s t}),
\end{equation}
where exponents and amplitudes have been expanded to first order in $x, \epsilon_1,$ and $\epsilon_2$.  Within this approximation, deviations from the assumed exponential decay will be equal and opposite for $\delta=\pm 1$ and will tend to cancel when the two cases are averaged.  In practice we check both extremes for consistency and take the average.

Experimentally, population remaining in $S_{1/2}$ is determined by shelving the population to $D_{5/2}$ and measuring the population scattered into $D_{3/2}$.  Shelving errors influence the initial amplitude of the exponential decay, and switching transients for the probe give rise to an initial time-dependent scattering rate.  Ignoring higher order corrections, the inferred $S_{1/2}$ population will have the form
\begin{equation}
\label{Eq:ShelvingError}
P(t)=(1-\delta s)\exp\left(-\int_0^t \gamma(t') dt' \right),
\end{equation}
where $\delta s$ determines shelving errors and $\gamma(t)$ describes the time-dependent scattering rate due to switching transients.  Taking the ratio of $P(t)$ measured at times $t_0$ and $t_1$ gives
\begin{equation}
\frac{P(t_1)}{P(t_0)}=\exp\left(-\int_{t_0}^{t_1}\gamma(t')d t'\right)=e^{-\gamma_s(t_1-t_0)},
\end{equation}
where we have assumed any switching transients occur for $t<t_0$.   
\section{Matrix element determination}
\subsection{Experimental system}
\begin{figure}[!ht]
  \centering
  \includegraphics[width=0.8\linewidth]{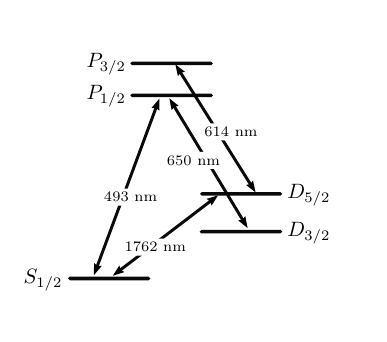}
  \caption{\label{fig:BaLevels}Ba$^+$ level structure showing the transitions used in this work.}
\end{figure}

The relevant level structure of Ba$^+$ is shown in Fig.~\ref{fig:BaLevels}.  Doppler cooling is achieved using near resonant light at 493\,nm and 650\,nm with light collected at 650\,nm used for detection.  State selective detection is achieved by shelving population in $S_{1/2}$ to $D_{5/2}$ using a clock laser at 1762\,nm.  Population transferred to $D_{5/2}$ is cleared by a repump laser at 614\,nm.  The experimental setup is identical to that described in \cite{jayjong2026zero}, with the Stark-shifting beam in that work replaced by the detuned probe at $493\,\mathrm{nm}$ used here for the matrix element determination.  

\subsection{The 493\,nm probe laser}
The detuned probe at $493\,\mathrm{nm}$ for the matrix element determination is generated via second-harmonic generation in a resonant frequency-doubling cavity.  The $986\,\mathrm{nm}$ fundamental is locked to a cesium-referenced transfer cavity, and the frequency is monitored in real time by a beat note against an optical frequency comb (OFC).  The beat note is monitored in one minute intervals, which allows drifts of $< 2\,\mathrm{MHz}$ in the transfer-cavity setup to be tracked throughout the measurements.  Similar to previous work, the 493\,nm probe propagates at an angle of approximately $68^\circ$ to an applied static magnetic field and is nominally linearly polarized orthogonal to the field.

Detuning of the probe beam is with respect to the resonant frequency of the $S_{1/2} – P_{1/2}$ transition, which was determined by measuring the scattering rate out of $S_{1/2}$ as a function of frequency and locating the peak of the resulting spectral feature.  Population was optically pumped into the $S_{1/2}$ state, a weak near-resonance beam at $\Delta\approx \pm \Gamma/2$ then scattered approximately 50\% of the population to $D_{3/2}$, and a servo was used on the laser offset to balance the populations on each side, thereby determining the center of the line.  This gave a resonant frequency of
\begin{equation}
f_{0}(S_{1/2} \,\textendash\, P_{1/2}) = 607\,426\,262\,645.5(7.1)\,\mathrm{kHz},
\end{equation}
where the uncertainty is purely statistical.  The measurement was carried out in 2021, with the probe propagating along the direction of a 0.2\,mT magnetic field with linear polarization.  In this configuration the most important systematic is in the population imbalance in the $m_J=\pm1/2$ states after Doppler cooling and optically pumping into $S_{1/2}$.  From simulations using the measured 48/52\% population imbalance, we estimate a systematic shift of less than 100\,kHz, which can be taken as an uncertainty.  The value is in good agreement with the 607\,426\,262.5(2)\,MHz given in \cite{dijck2015determination}, which was found in a less direct way.  The difference in the two values has no bearing on the final analysis as the associated uncertainty is well below the measurement precision.

\begin{figure}[!ht]
  \centering
  \includegraphics[width=0.8\linewidth]{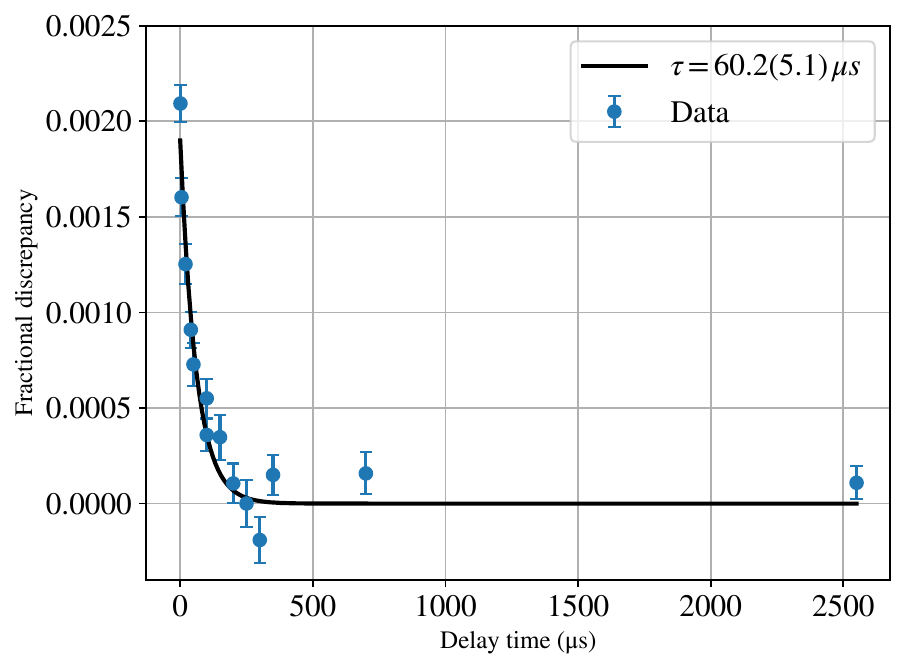}
  \caption{Transient response of the Stark AOM, plotted as the fractional discrepancy of the Stark shift at time $t$ relative to the reference shift at $t_{\text{ref}} = 5.0\,\mathrm{ms}$. An exponential decay fit is shown to illustrate the observed trend, rather than to serve as a physical model of the dependence.}
  \label{fig:Stark_AOM_respond}
\end{figure}

Power of the probe is monitored after passing through the vacuum chamber using a photodetector with a fixed 160\,MHz bandwidth.  The signal is used to control the driving power of a switching AOM, bringing the beam power to $>99\%$ of its set point within $10\,\mu\mathrm{s}$ with a slower convergence to the final set point. To characterize this slower time constant, the Stark shift induced by the probe was measured at various time delays after switching and compared against a reference shift at $t_\text{ref} = 5\,\mathrm{ms}$, which is long enough to ensure the laser intensity has fully stabilized.   Results are shown in Fig.~\ref{fig:Stark_AOM_respond}.  Based on these measurements, a $200\,\mathrm{\mu s}$ delay is used to measure Stark shifts to ensure minimal influence in the matrix element determination.

As noted in the previous section, it is advantageous to minimize the vector shift induced by the probe to suppress any imbalance in the circular components of the probe beam, which can otherwise influence the decay.  This is achieved by using a quarter wave plate after a linear polarizer to compensate any birefringence from the vacuum window. The vector shift is extracted by measuring the Stark shifts between the $\ket{S_{1/2}, \pm 1/2}$ and $\ket{D_{5/2}, \pm M_J}$ Zeeman sublevels, where $M_J \in \{1/2, 3/2, 5/2\}$. Because the vector component of the Stark shift changes sign with $M_J$ while the scalar and tensor components do not, the measured shifts can be decomposed to isolate the vector contribution. The QWP angle is adjusted until this extracted vector shift is minimized, as shown in Table \ref{tab:vector_shift_qwp}.  From the approximate 470\,kHz scalar shift, we infer $\epsilon$ given in Eq.~\ref{Eq:imbalance} to be $\sim 10^{-4}$ at the minimum value given in the table.  Repeated measurements throughout the matrix element determination did not see significant variation from this setting.

\begin{table}[!ht]
\centering
\caption{Differential vector shift induced by the scattering beam as a function of the quarter wave plate angle with respect to the zero of the vernier scale on the adjustable waveplate mount. Uncertainties are given in parentheses.  For comparison, the corresponding scalar shift was approximately 470\,kHz.}
\label{tab:vector_shift_qwp}
\begin{ruledtabular}
\begin{tabular}{@{\hspace{1cm}} d @{\hspace{1cm}} D{.}{.}{1.4} @{\hspace{1cm}}}
\multicolumn{1}{c}{Angle (deg)} & \multicolumn{1}{c}{Vector shift (kHz)}{\hspace{0.5cm}} \\
\hline
-1.00 & -17.246(45) \\
+0.50  & +0.065(39) \\
+1.00  & +8.199(45) \\
+2.00  & +18.851(47)
\end{tabular}
\end{ruledtabular}
\end{table}

\subsection{Measurements}
Measurements of Stark shifts and scattering rates are carried out in an interleaved fashion.  Each cycle begins with Stark-shift ($\Delta_s$) measurements on the $\lvert S_{1/2}, m=\pm 1/2 \rangle - \lvert D_{5/2}, m'=\pm 1/2 \rangle$ clock transitions. Each Zeeman line is probed on either side of its resonance with and without the probe beam present, giving eight measurements in total.  Subsequently, scattering-rate ($\gamma_s$) measurements are performed at two durations, $t_0 = 800\,\mu\mathrm{s}$ and $t_1 = 4\,\mathrm{ms}$, chosen to yield ground-state populations of approximately $0.8$ and $0.3$, respectively.  The two scattering rate measurements are performed starting with population pumped into $\lvert S_{1/2}, -1/2 \rangle$ and repeated with population pumped to $\lvert S_{1/2}, 1/2 \rangle$, to check for any dependence on population distribution.  These four scattering measurements are repeated three times within the cycle.  Each cycle is repeated 100 times before updating a servo that tracks the Stark shifts and magnetic field. 

\begin{figure*}
  \centering
  \includegraphics[width=0.95\linewidth]{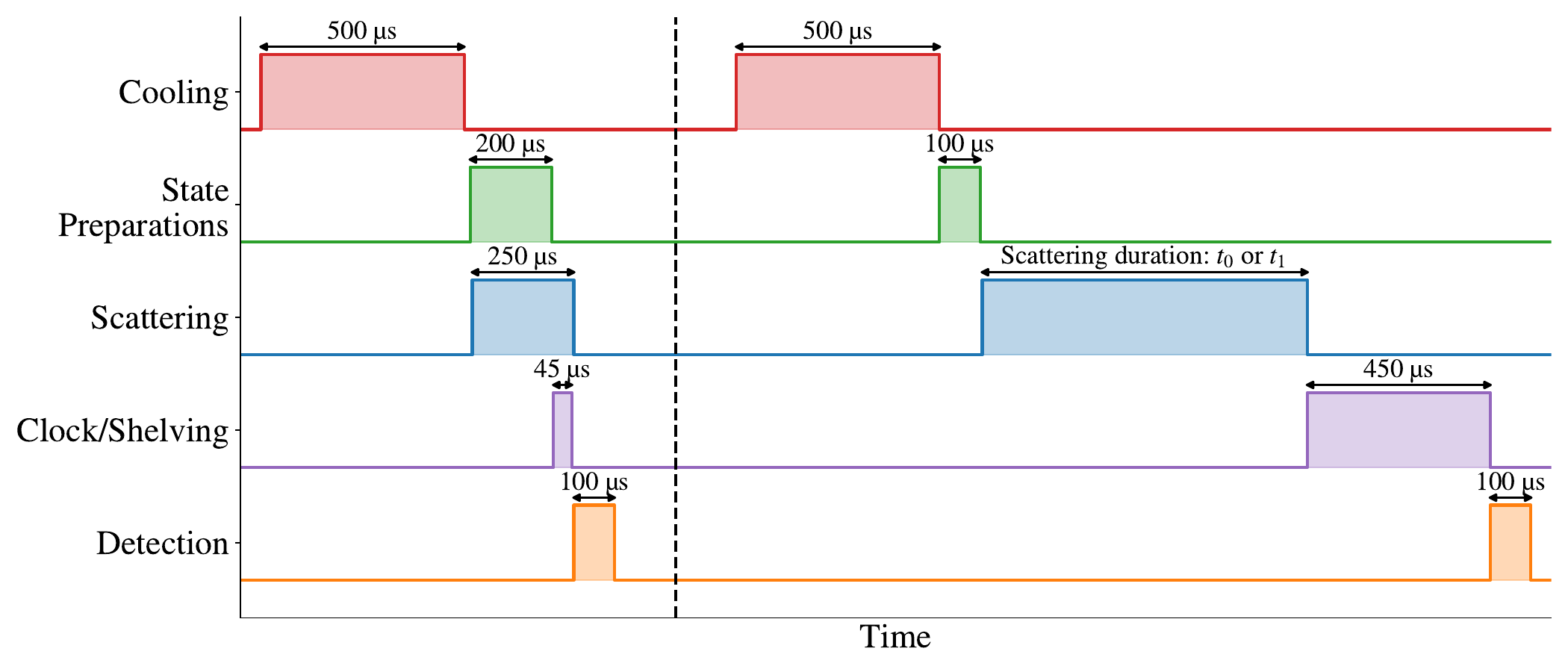}
\caption{\label{fig:time_sequences} Timing sequences for Stark shift measurements (left) and scattering measurements (right).  Clock pulse for the scattering measurements represents that total time for the six separate pulses discussed in the text.}
\end{figure*}

The timing sequences for the Stark-shift and scattering rate measurements within a single cycle are shown in Fig.~\ref{fig:time_sequences}.  Either measurement starts with Doppler cooling for $500\,\mathrm{\mu s}$ driving the transitions at 493\,nm and 650\,nm with repumping at 614\,nm to clear any population in $D_{5/2}$.  This is followed by state preparation to optically pump population into $\lvert S_{1/2},\pm 1/2 \rangle$.   For Stark shift measurements, the probe laser is switched on during state preparation, which has an increased duration relative to scattering rate measurements.  The probe laser does not influence state preparation and the increased duration allows the probe laser intensity to fully settle to its set point as shown in Fig.~\ref{fig:Stark_AOM_respond}.  For scattering rate measurements, measurement of population scattered to $D_{3/2}$ is achieved by shelving $S_{1/2}$ population to $D_{5/2}$ using the clock laser, which drives all six $\lvert S_{1/2}, \pm 1/2 \rangle - \lvert D_{5/2}, m'\rangle$ transitions sequentially to maximize the population transfer.

Matrix element determinations were carried out at $\Delta\approx\pm 2\pi\times 43\,\mathrm{GHz}$, and  $\Delta\approx\pm 2\pi\times 70\,\mathrm{GHz}$.  These values allow reasonable scattering rates for the available power, are sufficiently large to ensure Zeeman splittings and linewidth considerations are negligible, and still much smaller than the fine-structure splitting so that the 455\,nm transition has negligible contribution in the average.  At each detuning, the probe power was set to give a scattering rate of approximately $300\,\mathrm{s}^{-1}$, which is large enough to be accurately determined in a reasonable averaging time, but small enough that it does not degrade the lineshape of the clock transition when measuring the Stark shift.  

\begin{figure}
  \centering
  \includegraphics[width=0.95\linewidth]{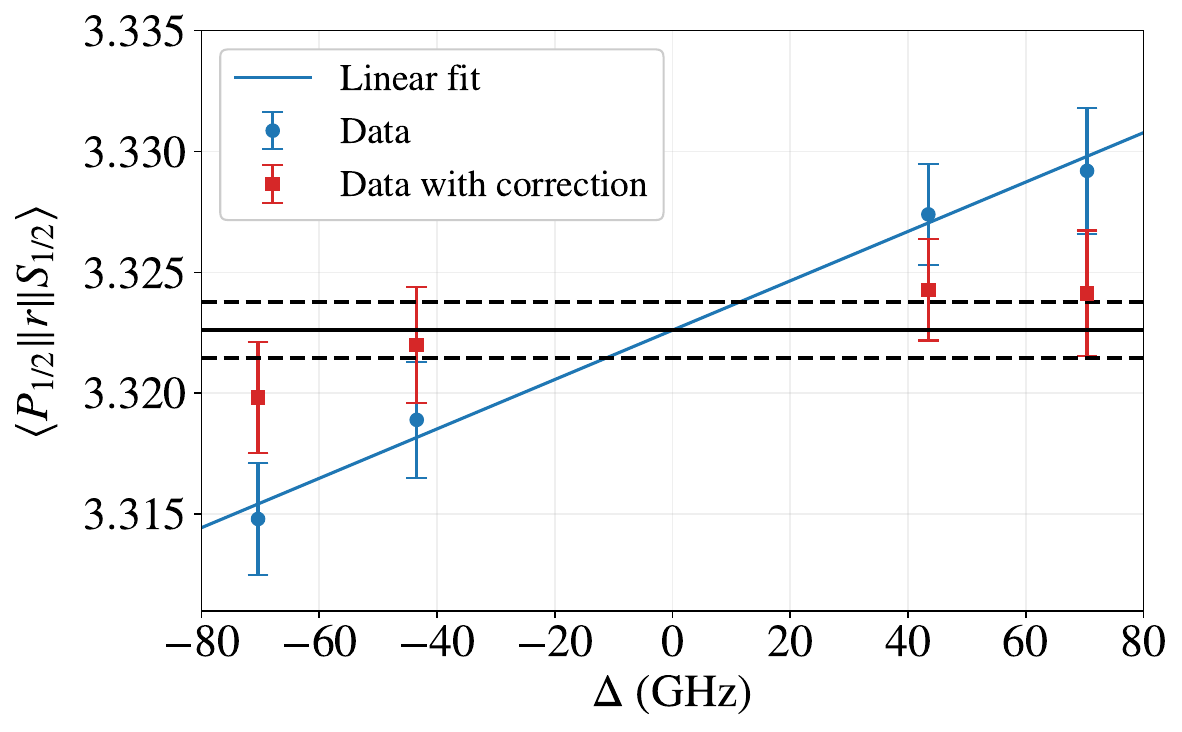}
    \caption{Measured matrix element $\redmel{P_{1/2}}{r}{S_{1/2}}$ as a function of $\Delta$. The raw experimental measurements are shown as blue circles and the data corrected for Stark-shift contributions from other levels are shown as red squares. Error bars on all data points represent $\pm 1\sigma$ uncertainties. The solid blue line represents a linear fit to the raw data. The fitted $y$-intercept, $3.322\,7(12)\,ea_0$, yields the matrix element value that cancels leading order corrections to the Stark shift given by Eq.~\ref{Eq:BasicSS}. The horizontal solid line indicates the overall average value, while the surrounding dashed lines mark the corresponding $\pm 1\sigma$ experimental statistical uncertainty.}
    \label{fig:rme_linear_fit}
\end{figure}

At each detuning there are approximately 1200 update cycles representing approximately 4 hours of averaging.  For each detuning the RME is calculated at each update cycle using $\Delta$ tracked by the frequency comb.  The average values and statistical uncertainty at each detuning are illustrated in Fig.~\ref{fig:rme_linear_fit} along with values that are corrected for Stark shift contributions beyond that given by Eq.~\ref{Eq:BasicSS}.  Uncorrected values are also tabulated in Table~\ref{tab:rme_detuning_avg} along with the averages over the individual detunings and the total average.  The detunings given in Table~\ref{tab:rme_detuning_avg} give the mean and half the difference of the maximum and minimum values over the full duration of the data collected for that configuration.  The overall average of $3.322\,7(12)$ given in Table~\ref{tab:rme_detuning_avg} is exactly the same as that determined by a straight line fit to the uncorrected data or the weighted average of the corrected data.

The values given in Table~\ref{tab:rme_detuning_avg} and Fig.~\ref{fig:rme_linear_fit} include the uncertainty from the measured value of $p$ \cite{arnold2019measurements}, which was reported as $p=0.268\,177\pm(37)_\mathrm{stat}-(20)_\mathrm{sys}$.  The systematic error in $p$ arose from dead-time considerations resulting in a bias to the overall value.  Here this bias is accounted for by taking $p=0.268\,167(47)$, which offsets $p$ by half the bias and increases the statistical uncertainty by the same.  Since the statistical uncertainty from the measurements of the RME dominate, we treat the uncertainty from $p$ as statistical, although this makes very little difference to the stated RME.  Treating the systematic uncertainty in $p$ separately, changes the RME by 1 in the last significant digit, which is captured by the increased statistical uncertainty in $p$.  

\begin{table}[htbp]
\centering
\caption{\label{tab:rme_detuning_avg}Measured $\redmel{P_{1/2}}{r}{S_{1/2}}$ values for different Stark-beam detunings and the averages obtained from pairs of opposite detunings. The final result is determined from the average of the measurements at $\pm 43\,\mathrm{GHz}$ and $\pm 70\,\mathrm{GHz}$. The detunings and their uncertainties are given by the average and half the difference of the maximum and minimum values recorded over the full duration of the experiment for that configuration.}
\begin{ruledtabular}
\begin{tabular}{d @{\hspace{1cm}} D{.}{.}{1.4} @{\hspace{1cm}}}
\multicolumn{1}{c}{Detuning (GHz)} & \multicolumn{1}{c}{$\redmel{P_{1/2}}{r}{S_{1/2}}{\hspace{0.5cm}} $} \\
\colrule
-70.391\,77(46) & 3.314\,9(23) \\
+70.393\,82(180) & 3.329\,3(26) \\
\multicolumn{1}{l}{Average:} & 3.322\,1(17) \\
\colrule
-43.442\,12(78) & 3.319\,0(24) \\
+43.442\,94(104) & 3.327\,5(21) \\
\multicolumn{1}{l}{Average:} & 3.323\,2(16) \\
\colrule
\multicolumn{1}{l}{Overall average:} & 3.322\,7(12) \\
\end{tabular}
\end{ruledtabular}
\end{table}

\subsection{Systematics}
Agreement between the individually averaged results and the straight-line fit to the uncorrected data is a mathematical requirement when uncertainties are equal. Agreement with the average of the corrected data reflects the fact that other systematic shifts are fractionally $< 2\times10^{-5}$.  Systematics include coupling to other levels, polarization dependent effects, Zeeman splittings, transients and leakage light from optical switching, and finite lifetimes of the metastable states.  As discussed in section~\ref{Methodology}, many of these are suppressed by the measurement protocol.  

A summary of the contributions to the fractional uncertainty in $\redmel{P_{1/2}}{r}{S_{1/2}}$ is given in Table~\ref{tab:error_budget}.  The uncertainty is dominated by the statistical uncertainty of the fit used in Fig.~\ref{fig:rme_linear_fit} and  This only changes final value of the matrix element by 1 in the least significant digit given.  All other considerations have no significant contribution to the final uncertainty as we now discuss.

\begin{table}[htbp]
\centering
\caption{\label{tab:error_budget}Error budget for the reduced matrix element
$\redmel{P_{1/2}}{r}{S_{1/2}}$. Entries are fractional contributions to the
RME, in units of $10^{-4}$. The statistical uncertainty is determiend from the straight line fit in Fig.~\ref{fig:rme_linear_fit}.  Systematic values are discussed in the text.}
\begin{ruledtabular}
\begin{tabular}{l d}
\multicolumn{1}{c}{Source} & \multicolumn{1}{c}{Fractional ($\times10^{-4}$)} \\
\colrule
Statistical                         & 3.6 \\
Leakage light (finite extinction)           & <0.2 \\
Vector shift / polarization imbalance       & <0.2 \\
Zeeman shifts                               & <0.2 \\
Detuning imbalance (equal \& opposite)         & <0.1 \\
Switching transients                        & <0.1 \\
Other-level / RWA corrections               & <0.01 \\
\colrule
Total & 3.6
\end{tabular}
\end{ruledtabular}
\end{table}

Coupling to other levels primarily results in corrections to Eq.~\ref{Eq:BasicSS}, primarily from the $P_{3/2}$ level.  Averaging over equal and opposite detunings suppresses this contribution to the order of $(\Delta/\omega_{FS})^2<2\times 10^{-6}$ fractionally where $\omega_{FS}$ is the fine structure splitting.  A slight imbalance in detunings of opposite sign degrades the suppression but it is still estimated to be below $10^{-5}$ for the worst case imbalances of a few MHz seen in the experiments.  Coupling to $P_{3/2}$ also contributes fractionally to the scattering rate at the same $(\Delta/\omega_{FS})^2$ level.

\begin{figure}
  \centering
  \includegraphics[width=0.95\linewidth]{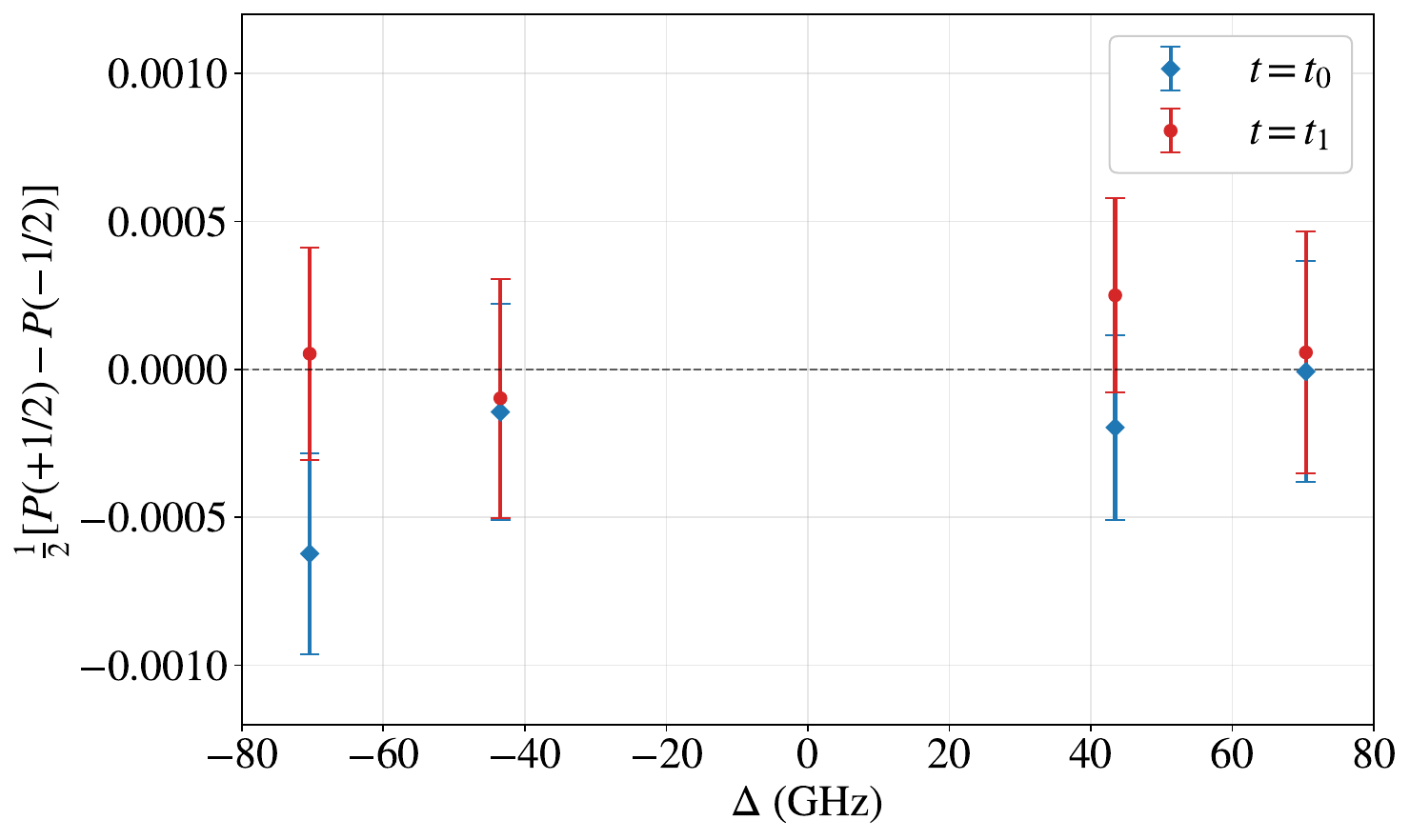}
    \caption{ \label{fig:OPeffect} Half the measured population differences when scattering from either $m$-state of $S_{1/2}$ at both $t_0$ and $t_1$ at each of the four detunings used in the measurements.  The statistical agreement with zero supports the negligible effects of residual deviations from linear polarization or Zeeman shifts.}
\end{figure}

Zeeman shifts and deviations from linear polarization result in non-exponential decay of the population as captured by Eq.~\ref{Eq:scattering}.  Measured suppression of the vector Stark shift gives $\epsilon/(k-1)\approx 7.6\times 10^{-5}$.  At $t_0$ this would give a fractional change of $\pm2\times 10^{-5}$ from the assumed exponential decay. This is also suppressed by averaging results with $\delta=\pm 1$.  In Fig.~\ref{fig:OPeffect} we give half the difference of the measured populations for $\delta=\pm1$ at both $t_0$ and $t_1$.   The statistical agreement with zero for both times supports the negligible effects of residual deviations from linear polarization or Zeeman shifts.  

Transients when switching the probe beam on can also result in non-exponential decay and the ratio of the populations at two times $t_0$ and $t_1$ mitigates this effect in addition to imperfect shelving.  As a consistency check, the scattering rates can be independently determined from the populations measured at either $t_0$ or $t_1$ alone by neglecting any shelving errors and assuming a simple exponential decay from which we infer RME estimates of $3.323\,4(15)$ or $3.323\,7(9)$, respectively.  These are statistically consistent, indicating that transient effects and shelving errors do not significantly affect the measured scattering rate.  The uncertainty when using measurements at $t_1$ alone are slightly smaller than that using the ratio, which arises from the nonlinear dependence on the measured populations.  Nevertheless, the scattering rate extracted from the population ratio was adopted for the final analysis, as this method suppresses systematic effects from transients and shelving imperfections.
 
A technical systematic that is not mitigated by the measurement protocol is leakage light from the cooling and optical pumping beams, which would contribute to measured scattering rates.  To quantify this effect, the cooling beams were nominally switched off, and the population remaining in the ground state was measured as a function of time. The leakage-induced scattering rates from the $\ket{S_{1/2}, +1/2}$ and $\ket{S_{1/2}, -1/2}$ states were measured independently and found to be statistically consistent and we therefore report the average value of $\gamma_s = 0.0103(36)\,\mathrm{s^{-1}}$ for a leakage-induced scattering rate. For comparison, the scattering rate produced by the 493~nm beam when the AOM is fully driven is $1.001(24) \times 10^{6}\,\mathrm{s^{-1}}$, corresponding to an effective extinction ratio of approximately $80\,\mathrm{dB}$. This is in agreement with independent measurements using a single-photon counting module (SPCM), which confirm an extinction ratio exceeding $70\,\mathrm{dB}$.  This was achieved by calibrating the input beam at high intensity with a power meter, and then determining the power in the off state with the SPCM, which avoids limitations from the dynamic range of the SPCM.  For the typical $300\,\mathrm{s}^{-1}$ used in the measurements, the residual scattering rate would only change the measured matrix element fractionally by at most $\approx 2 \times 10^{-5}$.

When optically pumping or shelving to a metastable state, the decay rate of that state can be a limitation.  The $D_{3/2}$ and $D_{5/2}$ levels have lifetimes of 80\,s and 30\,s respectively.  Hence decays of these levels during shelving or scattering with the probe will only influence measured populations at the $10^{-5}$ level and have no influence on the RME at the level of precision given.  We have also checked that leakage light at 650\,nm and 614\,nm, which can change the effective lifetimes, does not significantly alter this conclusion.
\section{Summary}
\begin{figure}[!ht]
  \centering
  \includegraphics[width=0.9\linewidth]{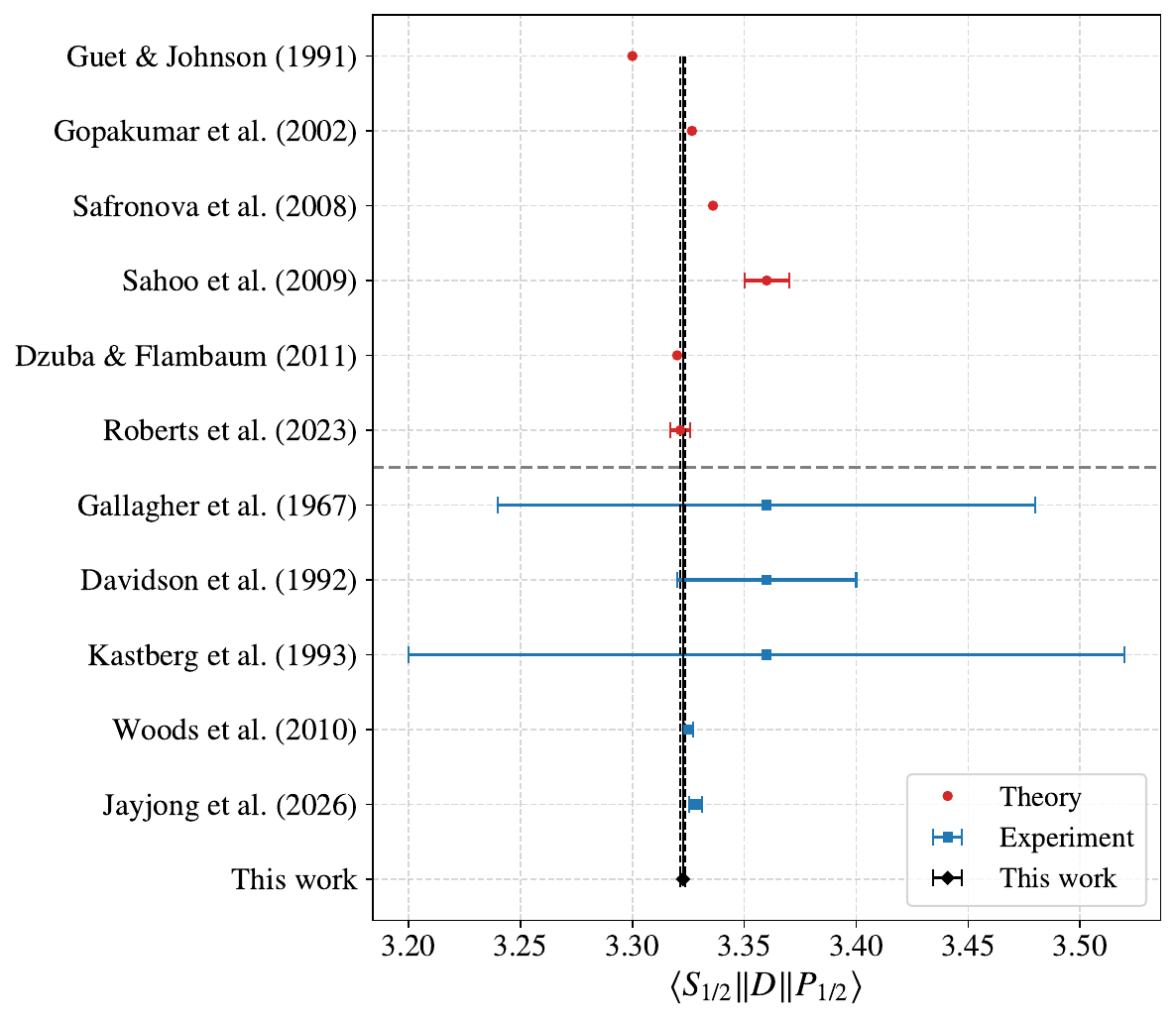}
\caption{  \label{fig:rme_results_comparision}
Comparison of the reduced electric-dipole matrix element
$\redmel{P_{1/2}}{r}{S_{1/2}}$ measured in this work with previous literature theoretical and experimental results.
Theoretical values are taken from Refs.~\cite{guet1991relativistic,gopakumar2002electric,iskrenova2008theoretical,sahoo2009light,dzuba2011calculation,roberts2023electric},
while experimental results are from Refs.~\cite{gallagher1967oscillator,davidson1992oscillator,kastberg1993measurements,woods2010dipole,jayjong2026zero}.
Uncertainties are shown in parentheses where available. In the figure, theoretical results are shown as circles and experimental results as squares, with horizontal error bars indicating reported uncertainties. The present work is represented by a diamond marker, with the central value shown as a solid vertical line and its $1\sigma$ uncertainty indicated by dashed lines.
}
\end{figure}

We have presented a high-precision measurement of the reduced electric-dipole matrix element $\redmel{P_{1/2}}{r}{S_{1/2}}$ in $^{138}\mathrm{Ba}^+$, obtaining a value of
\begin{equation}
\redmel{P_{1/2}}{r}{S_{1/2}}=3.322\,7(12).
\end{equation}
A comparison with previous theoretical and experimental results taken from the literature is shown in Fig.~\ref{fig:rme_results_comparision}.  Uncertainties in theoretical values are given where available.  The value we have given provides a factor of two improvement over the value reported in \cite{woods2010dipole}, which was obtained by the resonant excitation Stark ionization spectroscopy method, and the two values are within statistical agreement ($\approx 1\sigma$).

Taking $\redmel{P_{1/2}}{r}{S_{1/2}}=3.322\,7(12)$, the ratio reported in \cite{jayjong2026zero}, and branching fractions reported in \cite{arnold2019measurements,zhang2020branching}, provides fully experimental determinations of RMEs for all allowed electric dipole transitions in Fig.~\ref{fig:BaLevels}, and the lifetimes of the $P_{1/2}$ and $P_{3/2}$ levels, which are independent of the experimental results in \cite{woods2010dipole}.  We obtain matrix elements
\begin{subequations}
\begin{align}
\redmel{P_{1/2}}{r}{S_{1/2}}&=3.322\,7(12)\\
\redmel{P_{1/2}}{r}{D_{3/2}}&=3.039\,0(12)\\
\redmel{P_{3/2}}{r}{S_{1/2}}&=4.691\,0(17)\label{RME3}\\
\redmel{P_{3/2}}{r}{D_{5/2}}&=4.093\,4(17)\\
\redmel{P_{3/2}}{r}{D_{3/2}}&=1.328\,96(76)
\end{align}
\end{subequations}
which yields natural linewidths
\begin{align}
\Gamma_{3/2}&=2\pi\times 25.303(19)\,\mathrm{MHz}\\
\Gamma_{1/2}&=2\pi\times 20.232(14)\,\mathrm{MHz}
\end{align}
for $P_{3/2}$ and $P_{1/2}$ respectively with corresponding lifetimes $\tau_{3/2}=6.290\,0(47)$~ns and $\tau_{1/2}=7.866\,3(56)$~ns.

This work also provides an update to the estimate of $\Delta\alpha_0(0)=-73.09(12)$~a.u.\ for the static differential scalar polarizability of the $S_{1/2} - D_{5/2}$ clock transition.

The value of $\redmel{P_{3/2}}{r}{S_{1/2}}=4.691\,0(17)$ above has a $3\sigma$ discrepancy with the value of $4.701\,7(27)$ given in \cite{woods2010dipole}.  This arises from the significant discrepancy in the ratio $R_0=\redmel{P_{3/2}}{r}{S_{1/2}}/\redmel{P_{1/2}}{r}{S_{1/2}}$ reported in \cite{woods2010dipole} and \cite{jayjong2026zero}. The discrepancy is also the source of the difference in the matrix elements and $\Delta\alpha_0(0)$ reported in \cite{jayjong2026zero} as those values used results reported in \cite{woods2010dipole} in addition to the measured $R_0$.  Compared to the approach used in \cite{woods2010dipole}, branching fractions, zero-crossings of the clock transition's differential polarizability, and the matrix element determination presented here, are relatively free of systematics that could significantly alter the final result.  Nevertheless, it would be desirable to have an independent, direct measurement of $\redmel{P_{3/2}}{r}{S_{1/2}}$ to resolve the discrepancy and provide a consistency check on the results presented here.  The methodology we have demonstrated would be directly applicable to $\redmel{P_{3/2}}{r}{S_{1/2}}$ as well and should be able to obtain a similar level of uncertainty to resolve the difference.

\acknowledgements
This project is supported by the National Research Foundation, Singapore through the National Quantum Office, hosted in A*STAR, under its National Quantum Engineering Programme 3.0 Funding Initiative (W25Q3D0007) and under its Centre for Quantum Technologies Funding Initiative (S24Q2d0009).
\bibliography{ME493.bib}
\end{document}